\documentclass[12pt]{article}

\begin{document}
\leftskip10mm\rightskip10mm

\hfill {\bf SMC-PHYS-160}

\hfill {\bf hep-th/0001113}
\vskip 15mm
\centerline{\Large\bf An Early Proposal of ``Brane World"}
\vskip 20pt
\centerline{Keiichi Akama\footnote{
{\bf e-mail}: akama@saitama-med.ac.jp,\ \ 
{\bf telephone}: (81)492-95-1000 ext.\ 446,\ \ 
{\bf fax}: (81)492-95-4644.
}}
\centerline{Department of Physics, Saitama Medical College}
\centerline{Moroyama, Saitama, 350-0496, Japan}
\vskip 30pt

Here we place the \TeX-typeset version of the old preprint 
	SMC-PHYS-66 (1982),\footnote{
This cover is added in January, 2000.
The original preprint starts at the next page, and is identical to the 
	\vskip0pt\noindent\leftskip18pt
	published version except for minor typographical changes.
We place it on this preprint server, because,
	under the increasing interests and activities 
	concerning the brane world, 
	we often hear annoyance that the paper is hardly accessible in many of institutions,
	while most of them have internet access to the server.
We would like to thank Professor Ann~Nelson and Professor Matt~Visser
	for their useful suggestions in placing this old preprint on this preprint server.
}  
	which was published in
\vskip2mm\leftskip15mm\rightskip15mm\noindent
K. Akama, ``Pregeometry", in {\it Lecture Notes in Physics, 176, Gauge Theory and Gravitation, 
	Proceedings, 
	Nara, 1982}, 
	edited by K. Kikkawa, N. Nakanishi and H. Nariai, 
	(Springer-Verlag) 267--271.
\vskip2mm\leftskip10mm\rightskip10mm\noindent
In the paper, we presented the picture
	that we live in a ``brane world" (in the present-day terminology) {\it i.e.}
	in a dynamically localized 3-brane in a higher dimensional space.\footnote{
See also the related paper K.~Akama, Prog.\ Theor.\ Phys.\ {\bf 60} (1978) 1900, 
	where we show the bosonic and 
	\vskip0pt\noindent\leftskip18pt the fermionic brane-volume type actions 
	give rise to Einstein gravity on the brane through the quantum fluctuations.
}
We adopt, as an example, the dynamics of the Nielsen-Olesen vortex type 
	in six dimensional spacetime to localize our space-time within a 3-brane. 
At low energies, everything is trapped in the 3-brane, 
	and the Einstein gravity is induced through the fluctuations of the 3-brane.\footnote{
See also the related papers K.~Akama, Prog.\ Theor.\ Phys.\  
	{\bf 78} (1987) 184; {\bf 79} (1988) 1299; {\bf 80} (1988) 935, 
	\vskip0pt\noindent\leftskip18pt 
	which incorporates the normal connections of the brane as gauge fields.
} 
The idea is important because it provides a way basically distinct 
	from the ``compactification" to hide the extra dimensions
	which become necessary for various theoretical reasons.

\newpage
\leftskip0mm\rightskip0mm

\centerline{\large\bf PREGEOMETRY}
\vskip 12pt
\centerline{K. Akama}
\centerline{Department of Physics, Saitama Medical College}
\centerline{Moroyama, Saitama, 350-04, Japan}
\vskip 12pt

All the existing experimental evidences, though not so many, clearly support the general relativity
of Einstein as a theory of gravitation.
So far, extensive investigations have been made, based on the premise of general relativity.
Even the theory of induced gravity \cite{1}, or pregeometry, where the Einstein action is derived from
a more fundamental sage, are not free of this premise.
However, if the principle of the general relativity is true,
it should be a manifestation of some underlying dynamics.
just like the Kepler's law for the Newtonian gravity, or like the law of definite proportion 
in chemical reaction for atoms, etc.
So we would like to ask here why the physical laws are generally relative, instead we premise it.
The purpose of this talk is to propose a model to give possible answer to the question.
By general relativity, we mean general covariance of the physical laws in the curved spacetime.
Our solution, in short, is that it is because our four spacetime is a four-dimensional
vortex-like object in a higher-dimensional flat spacetime, where only the special relativity is assumed.
To be specific, we adopt the dynamics of the Nielsen-Olesen vortex \cite{2} in a six-dimensional
flat spacetime, and show that general relativity actually holds in the four-spacetime.
Furthermore we will show that the Einstein equation in the four-spacetime is effectively induced
through vacuum fluctuations, just as in Sakharov's pregeometry \cite{1}.

We start with the Higgs Lagrangian in a six dimensional flat spacetime
\begin{equation}
{\cal L}=-{1\over4}F_{MN}F^{MN}+D_M\phi^\dagger D^M\phi +a|\phi |^2-b|\phi |^4 +c
  \label{1}  \end{equation}
where 
$F_{MN}=\partial_MA_N-\partial_NA_M$ and $D_M \phi =\partial_M +ieA_M$.
This has the `vortex' solution \cite{2} 
\begin{equation}
A_M=\epsilon_{0123MN}A(r)X^N/r,\ \phi=\varphi(r)e^{in\theta},\ (r^2=(x^5)^2+(x^6)^2)
  \label{2}  \end{equation}
where $A(r)$ and $\varphi(r)$ are the solutions of the differential equations,
\begin{eqnarray}\displaystyle
&-\frac{1}{r}\frac{d}{dr}\left(r{d \over dr}\varphi\right)
+\left[\left({n \over r}+eA\right)^2-a+2b\varphi^2\right]\varphi=0\cr
&-{d \over dr}\left({1\over r}{d \over dr}rA\right)
+\varphi^2\left(e^2A^2+{en \over r}\right)=0
  \label{3}  \end{eqnarray}
The `vortex' is localized within the region of $O(\epsilon)$ $(\epsilon=1/\sqrt{a})$
in two of the space dimensions $(X^5,X^6)$, leaving a four-dimensional subspacetime
$(X^0-X^3)$ inside it. 
For large $a$, the curved `vortices' with curvature $R<<a$ become approximate solutions \cite{3},
which we denote by $A_M^0$ and $\phi^0$.
Let the center of the `vortex' be $X^M=Y^M(\xi^\mu)$ $(\mu=0-3)$, and 
take the curvilinear coordinate $x^M$ such that, near the `vortex',
\begin{equation}
X^M=Y^M(x^\mu)+n_m^M x^m,\ \ (M=0-3,5,6,\ \mu=0-3,\ m=5,6)
  \label{4}  \end{equation}
where $X^M$ is the Cartesian coordinate, and $n_m^M$ are the normal vectors of the `vortex'.
(Hereafter Greek suffices stand for 0-3, small Latin, 5, 6, and capital, 0-3, 5, 6).
Then the solution is 
\begin{equation}
A_M^0=\epsilon_{0123MN}A(r)x^N/r,\ \phi^0=\varphi(r)e^{in\theta}.\ (r^2=x^m x^m)
  \label{5}  \end{equation}

The S-matrix element between the states $\Psi_i$ and $\Psi_f$ is given by  
\begin{equation}
S_{fi}=\int\prod_{X^M}dA_Md\phi d\phi^\dagger
\exp\left[i\int{\cal L}d^6X \right]\Psi_f^*\Psi_i\prod_{X^M}\delta(\partial_MA^M)
  \label{6}  \end{equation}
We assume that the path integration is dominated by the field configurations 
of the approximate solutions (5) and small quantum fluctuations around it.
To estimate it, we first extract the collective coordinate by inserting 
\begin{equation}
1=\int\prod_{X_{/\!/}}dY^M(\xi^\mu)\delta\left(Y^M(\xi^\mu)-C^M(\xi^\mu)\right)
  \label{7}  \end{equation}
where $C^M(\xi^\mu)$ is the center of mass distribution of $|\widetilde\phi|^2$ 
$(\widetilde\phi=\phi-\sqrt{a/2b})$ in the normal plane $N(\xi^\mu)$ 
of the `vortex' at $x^\mu=\xi^\mu$,
\begin{equation}
C^M(\xi^\mu)=\int_{N(\xi^\mu)} X^M |\widetilde\phi|^2 d^2X_{\perp}\Bigg/
                 \int_{N(\xi^\mu)} |\widetilde\phi|^2 d^2X_{\perp}
  \label{8}  \end{equation}
By $\prod_{X_{/\!/}}$, we mean the product over four parameters $\xi^\mu$ 
with the invariant measure.
Then we transform them into the representation in the curvilinear coordinate $x^M$,
and we change the path-integration variables $A_M$ and $\phi$
to their quantum fluctuations $B_{\bar N}= A_{\bar N}-A^0_{\bar N}$ and $\sigma=\phi-\phi^0$,
retaining the terms up to quadratic in them.
\begin{equation}
S_{fi}=\int\prod_{X_{/\!/}}dY^M\prod_{X^M}dB_{\bar N} d\sigma d\sigma^\dagger
\delta(\sqrt{-g} \nabla_{\bar N}B^{\bar N}
\prod_{X_{/\!/}}\delta(\widetilde C^M)
\exp\left[i\int\left({\cal L}_0+{\cal L}_1 \right)\sqrt{-g}d^6x\right]\Psi_f^*\Psi_i
  \label{9}  \end{equation}
with 
\begin{eqnarray}
{\cal L}_0 &=& {\cal L}(\phi=\phi_0, A_M=A_M^0)
  \label{10}  
\\
{\cal L}_2 &=& 
-\frac{1}{2}g^{LM}\nabla_L B_{\bar N} \nabla_M B^{\bar N}
+B_{\bar N} B^{\bar N} e^2 |\phi^0|^2
+g^{LM}(D_L^0\sigma)^\dagger(D_M^0\sigma)\cr
&&-4ieV^{\bar N M} B_{\bar N} {\rm Im} \left( \sigma^\dagger D_M^0 \phi^0 \right)
+a|\sigma|^2 -b\left[ 4|\phi^0\sigma|^2 +2{\rm Re}(\sigma^\dagger\phi^0)^2\right],
  \label{11}  
\\
\widetilde C^m &=& \int x^m |\widetilde\phi|^2 dx^5 dx^6 \Bigg/ 
\int |\widetilde\phi|^2 dx^5 dx^6
  \label{12}  
\\
&=& \frac{1}{J_0}\int x^m \left[|\sigma|^2 + {\rm Re}(\widetilde \phi^0\sigma^\dagger)
\left\{1-\frac{2}{J_0}\int{\rm Re}(\widetilde \phi^0\sigma^\dagger)dx^5 dx^6\right\}\right]dx^5 dx^6,
  \label{13}  \end{eqnarray}
and $\widetilde C^\mu=0$, where the barred suffices stand for the local Lorentz frame indices,
$V^{\bar N M}$, the vierbein, $g^{LM}$, the metric tensor,
$\nabla_M$, the covariant differentiation, $D_M^0=\nabla_M+ieA_M^0$, 
and $J_0=\int|\widetilde\phi^0|^2dx^5dx^6$.
The Lagrangian ${\cal L}_2$ indicates that , outside the `vortex', any low energy fields are suppressed
because of the high barrier of $|\phi^0|^2$.
Inside the `vortex', $g_{m\mu}=O(R/a)<<1$, $g_{mn}=-\delta_{mn}+O(R/a)$ and 
$B_{\bar M}$ reduces to the four-vector $B_{\bar\mu}$ and two scalars $B_{\bar m}$.
Thus the spacetime looks like four-dimensional and curved to observers with large scale.
It is easily checked that the action is invariant under the general coordinate transformation
of the curved four-spacetime, i. e. the physical laws are generally relative!

Now  we see that the Einstein action is induced thorough vacuum polarizations.
The effective action $S^{\rm eff}$ for it is given by 
\begin{equation}
S^{\rm eff} = -i\ln\int\prod_{X^M}dB_{\bar N} d\sigma d\sigma^\dagger
\delta\left(\sqrt{-g}\nabla_{\bar N}B^{\bar N}\right)\prod_{X_{/\!/}}\delta(\widetilde C^M)
\exp\left[i\int\sqrt{-g}{\cal L}_2 d^6 x \right].
  \label{14}  \end{equation}
Exponentiating the argument of the $\delta$-functions by 
$\delta=\int dk e^{ikx}$, we get
\begin{equation}
S^{\rm eff} = -i\ln\int\prod_{\xi^\mu}dw_m\prod_{x^M}dB_{\bar M} d\sigma d\sigma^\dagger dv
\exp\left[i\int(\Xi\Phi+\Phi^\dagger \Delta\Phi)d^6x\right]
  \label{15}  \end{equation}
with 
\begin{eqnarray}
&&\hskip-5mm 
\Phi^\dagger=(B^{\bar M},\sigma,\sigma^\dagger),
  \label{16}  \\
&&\hskip-5mm  
\Xi=\sqrt{-g}(\nabla_{\bar M}v,
\ w_m x^m \widetilde\phi^{0\dagger}/J_0,
\ w_m x^m \widetilde\phi^0/J_0),
  \label{17}  \\
&&\hskip-5mm  
\Delta=\sqrt{-g}
  \label{18}  \\
&&\hskip-5mm 
\times\pmatrix{
\hskip-1mm\eta_{\bar M \bar N}\left(\frac{1}{2}\nabla_L\nabla^L+e^2|\phi^0|^2\right)
&
ieD_{\bar M}^0\phi^{0\dagger}
&
-ieD_{\bar M}^0\phi^{0}
\cr
-ieD_{\bar N}^0\phi^{0}
&
\hskip-3mm\frac{1}{2}D_L^0 D^{0L} +\frac{a}{2}-2b|\phi^0|^2+\delta_{11}^mw_m
&
-b(\phi^0)^2+\delta_{12}^m w_m
\cr
ieD_{\bar N}^0\phi^{0\dagger}
&
-b(\phi^0\dagger)^2+\delta_{21}^m w_m
&
\hskip-3mm\frac{1}{2}D_L^0 D^{0L} +\frac{a}{2}-2b|\phi^0|^2+\delta_{22}^mw_m
}
\nonumber
\end{eqnarray}
where $\delta^m$ is the nonlocal operator in 5-6 plane
\begin{equation}
\delta^m (x,x') = \frac{1}{2J_0}x^m\delta(x-x')
+\frac{1}{2{J_0}^2}(x^m+x'^m)
\pmatrix{\widetilde\phi^0(x)\cr\widetilde\phi^0(x)^\dagger}
\pmatrix{\widetilde\phi^0(x')^\dagger&\widetilde\phi^0(x')}
  \label{19}  \end{equation}
Performing the path-integration in $B_{\bar N}$, $\sigma$, $\sigma^\dagger$, 
and $v$, we get (with $\Xi_0=\Xi|_{v=0}$)
\begin{equation}
S^{\rm eff} = \frac{1}{2}i{\rm Tr}\ln\Delta
+\frac{1}{2}i{\rm Tr}\ln\left[\partial_M\sqrt{-g}(\Delta^{-1})^{MN}\sqrt{-g}\partial_N\right]
-\frac{1}{4}\int\Xi_0^\dagger\Delta^{-1}\Xi_0d^6x
  \label{20}  \end{equation}
$S^{\rm eff}$ in (\ref{20}) is estimated perturbatively in $h^{MN}=g^{MN}-\eta^{MN}$
($\eta^{MN}={\rm diag}(1,-1,-1,-1,-1,-1)$) and $w$.
The propagator is given by the inverse of $\Delta|_{h^{MN}=0,w=0}\equiv\Delta_0$.
$\Delta_0$ can be separated into two parts $\Delta_0^{\rm sp}$ and $\Delta_0^{\rm ex}$
which operates on four-space variables $x^\mu$, and the extra space variables $x^m$,
respectively.
Furthermore, these $\Delta_0$'s are block-diagonalized into two parts
$\Delta_0^{\rm V}$ and $\Delta_0^{\rm S}$, which operate on the four-vector $B^\mu$ 
and coupled scalars 
$(S^{(1)},S^{(2)},S^{(3)},S^{(4)})=(B^5,B^6,\sigma,\sigma^\dagger)$,
respectively. They are given by 
\begin{eqnarray}
&&\hskip-13mm 
\Delta_0^{\rm V,sp}=\frac{1}{2}\ \framebox(7,7){}\,,
\ \ \ 
\Delta_0^{\rm S,sp}=\frac{1}{2}\ \framebox(7,7){}\,,
\ \ \ 
\Delta_0^{\rm V,ex}=-\frac{1}{2}\partial_l \partial_l+e^2|\phi^0|^2,
\cr
&&\hskip-13mm  
\Delta_0^{\rm S,ex}
=\pmatrix{
\left(-\frac{1}{2}\partial_l\partial_l+e^2|\phi^0|^2\right)\eta_{mn}
&
ieD_{n}^0\phi^{0\dagger}
&
-ieD_{n}^0\phi^{0}
\cr
-ieD_{m}^0\phi^{0}
&
-\frac{1}{2}D_l^0 D_l^0 +\frac{a}{2}-b|\phi^0|^2
&
-b(\phi^0)^2
\cr
ieD_{m}^0\phi^{0\dagger}
&
-b(\phi^{0\dagger})^2
&
-\frac{1}{2}D_l^0 D_l^0 +\frac{a}{2}-b|\phi^0|^2
}
  \label{21}  \end{eqnarray}
where $\framebox(7,7){}\,= \eta^{\mu\nu}\partial_\mu\partial_\nu$.
Then the propagators for each class are given by
\begin{eqnarray}
\left[({\Delta_0^{\rm V}})^{-1}\right]^{\mu\nu}
&=&\eta^{\mu\nu}\sum_k(\,\framebox(7,7){}\,+{m_k}^2)^{-1}V_k(x^m)V_k(x'^m),
\cr
\left[({\Delta_0^{\rm V}})^{-1}\right]^{\mu\nu}
&=&\sum_k(\,\framebox(7,7){}\,+{m'_k}^2)^{-1}S_k^{(a)}(x^m)S_k^{(b)}(x'^m),
  \label{22}  \end{eqnarray}
where $V_k$, $S_k^{(0)}$, ${m_k}^2$ and ${m'_k}^2$ are the solutions and the eigenvalues 
of the differential equations in the extraspace,
\begin{equation}
\Delta_0^{\rm V,ex} V_k = {m_k}^2 V_k,
\ \ \ \ 
\Delta_0^{{\rm S,ex}(a)(b)} S_k^{(b)} = {m'_k}^2 S_k^{(a)}.
  \label{23}  \end{equation}

The argument of the logarithms in (\ref{20}) is expanded as follows
\begin{eqnarray}
&&\hskip-10mm
\Delta=\Delta_0(1+\Delta_0^{-1}\Delta_{\rm int}),
  \label{24}  \\
&&\hskip-10mm
\partial_M\sqrt{-g}(\Delta^{-1})^{MN}\sqrt{-g}\partial_N
=1+{\Delta'_0}^{-1}
+\partial_m(\Delta_0^{-1})^{mn}\partial_n+\Delta'_{\rm int},
  \label{25}  \end{eqnarray}
where $\Delta_{\rm int}$ and $\Delta'_{\rm int}$ are the interaction parts
including $h^{\mu\nu}$ and $w$, and
\begin{equation}
{\Delta'_0}^{-1}=\sum_k {m_k}^2(\,\framebox(7,7){}\,+{m_k}^2)^{-1}V_k(x^m)V_k(x'^m).
  \label{26}  \end{equation}
We expand the logarithms in (\ref{20}), and get series of one-loop diagrams
with external $h^{\mu\nu}$ and $w$ lines attached.
These diagrams diverge quartically in the ultraviolet region.
We introduce the momentum cutoff $\Lambda$ much larger than $\sqrt{a}$,
and calculate the divergent contributions. The diagrams with vertices which involve
extra-space operators are less divergent.

After this, the same argument as in the pregeometry \cite{1} leads to the Einstein
action in the four-dimensional curved space. 
Namely, the divergent contributions are
\begin{equation}
S^{\rm eff} = \int\sqrt{-g}\left[
(N_0\alpha_0+N_1\alpha_1+\alpha_c)\Lambda^4+
(N_0\beta_0+N_1\beta_1+\beta_c)\Lambda^2R\right]
d^4x
  \label{27}  \end{equation}
plus less divergent terms, where $N_0$ and $N_1$ are the numbers of the scalar
and vector bound-states in (\ref{23}), respectively,
and $\alpha_0$, $\alpha_1$, $\beta_0$, $\beta_1$ are calculable constants of $O(1)$.
The values are found in literatures \cite{1} and \cite{4},
though we should be careful, since they depend on the cutoff-method and even on gauge.
$\alpha_c$ and $\beta_c$ are the contributions from the continuum states in (\ref{23}).
Now, together with the contributions from ${\cal L}_0$, 
we finally get the Einstein action
\begin{equation}
S = \int\sqrt{-g}\left(\lambda + \frac{1}{16\pi G}R\right)d^4x
  \label{28}  \end{equation}
where
\begin{equation}
\lambda = \int{\cal L}_0 dx^5 dx^6 + (N_0\alpha_0+N_1\alpha_1+\alpha_c)\Lambda^4,
\ \ \ \ 
\frac{1}{16\pi G}=(N_0\beta_0+N_1\beta_1+\beta_c)\Lambda^2.
  \label{29}  \end{equation}

In conclusion, in this model:

\leftskip20pt
\vskip0pt\noindent\hskip-20pt 1)\vskip-15pt\noindent 
The principle of general relativity is induced, instead it is premised.

\vskip0pt\noindent\hskip-20pt 2)\vskip-15pt\noindent 
The Einstein equation is induced just as in Sakharov's pregeometry.

\vskip0pt\noindent\hskip-20pt 3)\vskip-15pt\noindent 
Two kinds of internal symmetries are induced, those of the transformation and
the excitation in the extra-space. 
The former is somewhat like isospin,
while the latter, generation.
This suggests a new mechanism for unification of the interactions.

\vskip0pt\noindent\hskip-20pt 4)\vskip-15pt\noindent 
When the gravitational field is quantized, 
the ultraviolet divergences should be cut off at the inverse of the size of the `vortex', 
which may be much smaller than the Planck mass.
If this is the case, we can by-pass the problem of renormalizability of the gravity.

\vskip0pt\noindent\hskip-20pt 5)\vskip-15pt\noindent 
Particles with sufficiently high energy can penetrate into the extra dimensions.

\vskip0pt\noindent\hskip-20pt 6)\vskip-15pt\noindent 
At very high temperatures \cite{5}, or high densities, 
the `vortex' is spread out over the extra-space 
revealing the higher dimensional spacetime.


\begin{thebibliography}{99}
\bibitem{1}
A.~D.~Sakharov,  Dokl.\ Akad.\ Nauk SSSR {\bf 177} (1967) 70;
Theor. Mate. Fiz. {\bf 23} (1975) 23;
K.~Akama, Y.~Chikashige, T.~Matsuki and H.~Terazawa,
	 {Prog.\ Theor.\ Phys.} {\bf 60} (1978) 868;
K.~Akama,  {Prog.\ Theor.\ Phys.} {\bf 60} (1978) 1900;
For a review, S.~L.~Adler, Rev.\ Mod.\ Phys.\ {\bf 54} (1982) 729. 

\bibitem{2}
H.~B.~Nielsen and P.~Olesen, Nucl.\ Phys.\ {\bf B61} (1973) 45.

\bibitem{3}
D. F\"orster, Nucl.\ Phys.\ {\bf B81} (1974) 84;
J.~L.~Gervais and B.~Sakita, Nucl.\ Phys.\ {\bf B91} (1975) 301.

\bibitem{4}
K.~Akama, Phys.\ Rev.\ {\bf D24} (1981) 3073.

\bibitem{5}
K.~Akama and H.~Terazawa, Gen.\ Relat.\ Grav.\ to be published.\footnote{
{\small\bf Note added in January, 2000:} 
K.~Akama and H.~Terazawa, Gen.\ Relat.\ Grav.\ {\bf 15} (1983) 201.
}

\end{thebibliography}
\end{document}